\documentclass[preprint,preprintnumbers,amsmath,amssymb,superscriptaddress]{revtex4}

\usepackage{graphicx}
\usepackage{dcolumn}
\usepackage{bm}

\makeatletter
\def\@seccntformat#1{}
\makeatother
\renewcommand{\numberline}[1]{}

\begin{document}
\title{Band Topology or Geometry?}
\author{Yi-Dong Wu}
\affiliation{Department of Applied Physics, Yanshan University, Qinhuangdao, Hebei, 066004, China}
\email{wuyidong@ysu.edu.cn}

\maketitle

\textbf{ The study of the topology of energy bands in solid has always been interesting and fruitful. Historically, Thouless et al proposed the TKNN number or Chern number of the energy band to explain the quantization of Hall conductance in the integer quantum Hall effect. Recently, $Z_2$ topological insulators have been intensively studied. There are gapless edge states in these materials which are protect by bulk band topology. The topology of the bands is characterized by some $Z_2$ invariants. However, $Z_2$ invariants are crude and strongly dependent on the symmetry. Here we give an unified picture of the relationship of the edge states and the geometry of the energy bands. We show the band geometry determines not only the topological but also the geometrical properties of the edge states. Our picture is applicable even when the symmetries are broken.}\\

To explain the  quantization of Hall conductance in the integer quantum Hall effect Thouless et al calculated the Hall conductance  of the two dimensional electron gas in a uniform magnetic field and period potential using the Kubo formula\cite{Thouless1982}. It's shown the quantization of Hall conductance is a bulk property and directly related to a the TKNN number or Chern number of the energy bands. However,the quantum Hall effect in a finite system with edge depends on the existence of the gapless edge state. Then Hatsugai established correspondence between Chern number and gapless edge states \cite{Hatsugai1993}. In 1988 Haldane proposed a tight-binding model which support a occupied energy band with nonzero Chern number in ground state\cite{Haldane1988}. The system can exhibit quantum Hall effect in the absence of macro magnetic field. We refer such materials as Chern insulators. The Chern insulators only exist when time reversal symmetry is broken. Some of the early proposed time-reversal symmetric topological insulators can be viewed as two copy of Chern insulators. In these insulators there are one component of the spin or pseudospin is conserved\cite{Qi2006,Onoda2005,Bernevig2006,Kane2005a}. So the occupied bands of the ground states can be easily divide to two bands with opposite nonzero Chern numbers according to the value the spin or pseudospin takes and there are clear energy band and edge state correspondence just like in the Chern insulators. In these cases the physics pictures are quite simple: the two time-reversal related bands  pump the spins or pseudospins to opposite directions when electric field present and the quantum spin Hall effect can be easily interpreted in this way. However when there are no conserved spin or pseudospin as in \cite{Kane2005b} when the Rashba term are nonzero, the two occupied bands don't naturally form a time-reversal related pair. So it's claimed the Chern numbers lost meaning and the $Z_2$ topological invariant is proposed to characterize the state\cite{Hasan2010}. The $Z_2$ classification is mathematically involved and make the physical picture complicated. It's even claimed the isolate $Z_2$ insulator will return to its origin state after two pumps\cite{Fu2006}. We will show the Chern number can still be used to classify the two dimensional topological insulators and the $Z_2$ topological insulator can be viewed as two Chern insulators put together\cite{Wu2011}. The three dimensional topological insulators are proposed and observed shortly after\cite{Fu2007,Zhang2009,Xia2009}. It's classified by four $Z_2$ invariants. The strong topological insulators have odd number of Dirac points at the surface. The existence of these $Z_2$ invariants depends on the time-reversal symmetry. With the similar idea a topological crystalline insulators are proposed with $C_4$ or $C_6$ symmetry and the $Z_2$ invariant characterize this insulator depends on these symmetries\cite{Fu2007}. \\

All those topological invariants are somewhat crude, that is they only determine the topology of the edge state. Historically the word ``topology" was first used to distinguish ``qualitative geometry from the ordinary geometry in which quantitative relations chiefly are treated"\cite{Guthrie1883}. The quantitative geometry certainly contains more details than the ``qualitative geometry". In this article we show how the geometrical property of the energy bands is studied and how it's related to the edge state property. The first problem is to distinguish individual energy band. In most of insulators there are several occupied energy bands, e.g. because of time-reversal symmetry ground state of two dimensional $Z_2$ insulator have even number of occupied bands. These bands form a trivial vector bundle. The topological non-triviality come from the time-reversal symmetry. We can always find global continuous bands that span a trivial bundle. However,if we insist on using two groups of time-reversal related bands to span the bundle, the two groups of bands won't be global continuous and each has an odd Chern number\cite{Wu2011}. We propose a natural way to define the two group of time-reversal related bands. In this way we can establish the energy band and edge state correspondence just like in Chern insulators. Let $\mathbf{a_1}$ and $\mathbf{a_2}$ denote the primitive vectors of the two dimension lattice. $\mathbf{a_1}$ parallel to the edge. The wave vectors of the occupied bands are functions of $n_1=\mathbf{k}\cdot\mathbf{a_1}$ and $n_2=\mathbf{k}\cdot\mathbf{a_2}$. For simplicity we assume two bands are occupied, all our results can be easily generated to more bands. Two global continuous orthonormal wave vectors $|u_1(n_1,n_2)\rangle$ and $|u_2(n_1,n_2)\rangle$ can be defined to span the occupied space. When $n_1$ is fixed the problem become one-dimensional. To make a natural decoupling of the occupied bands in one dimension we demand the in each of the decoupled bands the wave vectors are smoothly defined, that is they can be connected by parallel transport along $n_2$. Parallel transport is a geometrical property of the occupied bands and the geometrical structure of the occupied bands are determined by the non-Abelian Berry connection. From real space point of view the natural decoupling should make the one-dimensional Wannier functions as localized as possible. In fact the two approaches give the same result\cite{Kivelson1982}. A parallel transport of wave vectors $|u_i\rangle(i=1,2)$ at $n_2=n_{20}$ along $n_2$ can be expressed as
\begin{equation}
\left(|v_1(n_2)\rangle,|v_2(n_2)\rangle\right)=\left(|u_1(n_2)\rangle,|u_2(n_2)\rangle\right)P\exp\left(i\int_{n_{20}}^{n_2}A_{2}(n_2^{'})dn_2^{'}\right)
\end{equation}
where
\begin{equation}
A_{2}=\left(
        \begin{array}{cc}
          \langle u_1|\frac{\partial}{\partial n_2}|u_1\rangle  & \langle u_1|\frac{\partial}{\partial n_2}|u_2\rangle \\
          \langle u_2|\frac{\partial}{\partial n_2}|u_1\rangle & \langle u_2|\frac{\partial}{\partial n_2}|u_2\rangle \\
        \end{array}
      \right)
\end{equation}
is the non-Abelian Berry field, it's a $2\times 2$ matrix\cite{Bohm2003}. For notation convenience we drop the $n_1$ temporarily. $|v_1(n_2)\rangle,|v_2(n_2)\rangle$ are smooth along $n_2$. However there is a mismatch $U_g(n_{20})=P\exp\left(i\int_{n_{20}}^{n_{20}+2\pi}A_{2}(n_2^{'})dn_2^{'}\right)$ between $|v_i(n_{20}+2\pi)\rangle$ and the $|v_i(n_{20}\rangle=|u_i(n_{20}\rangle$. To cancel this mismatch we diagonalize $U(n_{20})$ by a $U(2)$ transformation. Then we perform the same $U(2)$ transformation at $|u_i\rangle$ for all $n_2$ and in this way we establish a new frame. It's easy to prove in the new frame there are only two $U(1)$ mismatches between $|v_i(n_{20}+2\pi)\rangle$ and the $|v_i(n_{20})\rangle=|u_i(n_{20})\rangle$, which are the eigenvalues of the $U_g(n_{20})$. In the new frame $|v_i(n_{2})\rangle$ can be connected by parallel transport except an $U(1)$ number and the occupied bands are naturally decoupled to two. The eigenvalues $\exp(i2\pi\Phi_{i})$ of $U_g(n_{20})$ are independent of $n_{20}$ and if they are nondegenerate the decoupling will be unique. For all $n_1$ we choose $n_{20}=0$ and make the decoupling. Here we only consider the case eigenvalues  of $U_g(n_{20})$ are degenerate for isolate $n_1$ and the decoupling at the degenerate points can be determined by continuous extension. In this way the two occupied bands of the two-dimensional system are naturally decoupled and the bands are smooth along the $n_2$ direction. All our discussions are based on such a decoupling which is certainly depend on the particular edge we study. So we don't study the general geometry of the bands but the geometry of bands for a given edge. $\Phi_{i}$, defined modulo $1$, give real space position of the center of the Wannier functions. $\Phi_{i}$ exhibit the geometry property of the bands. The change of $\Phi_{i}$ in one period of $n_1$($2\pi$) is the Chern number of the band. In the $Z_2$ insulator because of the time-reversal symmetry the decoupled bands are related by time-reversal operator and $\Phi_{1}(n_1)=\Phi_{2}(-n_1)$. The two bands have opposite Chern numbers and $Z_2$ is the parity of the Chern number. In \cite{Fidkowski2011} with a simplified model of edge topology of the edge state is identified with that of centers of the Wannier functions $\Phi_{i}(n_1)$. So we have the energy band edge state correspondence: each of the two dimensional naturally decoupled bands with nonzero Chern number determines a branch of gapless edge state. In two dimension insulator the study of topology of energy bands is reduced to study the topology of the lines formed by $\Phi_{i}$s. Same idea can be applied in three dimension. In three dimension the topology of the bands are determined by the topology of surfaces of $\Phi_{i}(n_1,n_2)$, where $n_1=\mathbf{k}\cdot\mathbf{a_1}$,$n_2=\mathbf{k}\cdot\mathbf{a_2}$ and $\mathbf{a_1}$, $\mathbf{a_2}$ are the primitive vectors at the two dimensional edge. However, in three dimension, because of the complexity of the topology of the surface, e.g. existence of the ``Dirac" type of points, the energy bands can't always be decoupled. The above discussions can be easily applied to the time-reversal symmetry broken materials with several occupied bands. In this work we study several types of topological insulators with and without symmetry protection by direct comparing the geometry of the $\Phi_{i}$ and the spectrums of edge states. We conclude that the geometry of the naturally decoupled bands determine not only the topology but also the geometry of the edge state. So the edge state of those topological insulators actually protect by the geometry the occupied bands.\\

Our first example is graphene, the first proposed $Z_2$ topological insulator\cite{Kane2005b}. In FIG 1 we compare the geometry of the $\Phi_{i}$s and that of the spectrum of edge states. We see edge state occurs at where the geometry of the bands is more nontrivial and there are no edge states at where $\Phi_{i}$s are relatively flat. The geometrical structure of the spectrum of the edge state and that of $\Phi_{i}$ are very similar. When the parameter $\lambda_v$ varies granphene experience a topological and trivial insulator phase transition. The $\Phi_{i}$ not only indicate the topological change but also accurately predict where the gap of edge state opens. Similar result can be obtained from the HgTe quantum well. We use the simple tight-bonding model in \cite{Bernevig2006} with the $d=70\mathbf{A}$. We can see $\Phi_{i}$s are quite flat except a small interval around the $n_1=0$ where the the edge state occur. We also calculated the $\Phi_{i}$s of a $Z_2$ trivial topological insulator in \cite{Qi2006}. The occupied bands are naturally decoupled to two time-reversal related bands with Chern number $\pm2$. Compare the $\Phi_{i}$s with the edge state spectrum we see the gapless edges states are actually protected by the geometry of the bands. In this case the degeneracies at the cross points of the edge state spectrum are very similar to the ``accident degeneracies" introduced by zone folding when cell are enlarged\cite{Fu2007b}. It's claimed these degeneracies aren't protect by symmetry so can be easily lift by random potentials lower the translational symmetry. However numerical simulation shows they are at lest quite strong against time-reversal symmetric random potentials along the $n_2$. Though the position of the cross points varies with different random potential the degeneracies aren't lifted by rather strong potentials. So whether the geometrical protected edge states in this case is trivial is still an open problem.\\

In Fig 2 we show some comparisons between the $\Phi_{i}$s as function of $n_1$ and $n_2$ which are surfaces and the spectrum of edge state of the tree dimensional topological insulator in \cite{Fu2007}(For convenience the we draw the energy spectrums which are most close to the bulk gap, only the part cross the bulk gap are actually are edge state). We see the conclusions from two dimensional case are still applicable here: edge states only occur when the geometry of $\Phi_{i}$s are more nontrivial. Moreover, in the cases of three  dimensional topological insulators our geometrical representation shows great advantage over the topological indices in \cite{Fu2007}. The geometrical representation predicts not only the existence of the edge state Dirac cones but also where the Dirac cone appears. So the topological indices only give part of the topological information that can be inferred for the geometrical structure of the surfaces of the $\Phi_{i}$s. Our geometrical representation clearly show the picture of three dimensional topological in \cite{Roy2009a,Roy2009b} fails. Their picture claim there must be monopole of Berry field for the Chern number change between different surfaces in Brillouin zone which means occupied bands must ``collide" with each other at the middle of half Brillouin zone. From Fig 2 we can easily observe that the differences of  Chern numbers of the bands in the two dimensional surface come from the switch partners at the Dirac type of points of $\Phi_{i}$s. In Fig 3 we show comparisons $\Phi_{i}$s and edge state spectrum in $Be_2Se_3$ with the tight-binding model in \cite{Li2010}. The correspondence are almost perfect.\\

In \cite{Fu2007} the so called topological crystalline insulators are proposed and it's claimed the gapless edge states in these materials are protected by the time reversal symmetry and the $C_4$ or $C_6$ rotation symmetry. A rather complicate $Z_2$ topological invariant is proposed to characterize the topology od the occupied bands and the invariant depend on both the time reversal symmetry and the $C_4$ or $C_6$ rotation symmetry. It's shown in Fig 4 the $\Phi_{i}$s not only indicated the topological property of edge state spectrum but also accurately predicted the geometrical structure of the edge state. We also deliberately break the $C_4$ rotation symmetry by using different parameter $t_{2}^{'}$ in $x$ and $y$ directions in the proposed tight-binding model. In this modified model there is only a $C_2$ rotation symmetry left, the discussions of the topological properties of the bands in \cite{Fu2007} are no longer applicable, that is, the $Z_2$ invariant can't be defined here. However, we shows the edge states geometry or topology aren't changed by the broken $C_4$ symmetry even with a very large difference of $t_{2}^{'}$s in $x$ and $y$ directions. In Fig 4 we see the $\Phi_{i}$s can still predict the edge state geometry when the symmetry is broken. So we propose in this case it's more convenient to use the geometry of the bands to characterize the insulator than to use some complicated topological invariant.\\

The study of the geometry of the bands becomes more useful when the time-reversal symmetry is broken. The time-reversal symmetry is either broken by external applied field or spontaneously as in \cite{Yang2011,Yu2010}. In these cases the $Z_2$ classification is certainly no longer applicable. Our natural decoupling of the bands doesn't depend on symmetry of the system, so the edge state properties can still be inferred from the study of the geometry of the bands. We first consider the case graphene with a uniform exchange field that acts on the $z$ component of spin of electron through Zeeman's type of coupling as in \cite{Yang2011}. In their work the edge states are characterized by the spin Chern number. Though their conclusion is qualitatively correct, the physics picture isn't clearly explained and it's somewhat unnatural to use spin Chern number when the ground state become ferromagnetic since the spin Chern number treats up and down spins equally. In fact the physics picture is quite simple and clear in this case. When the Rashba coupling is absent the Hamiltonian can be considered as two pieces of Haldane' Hamiltonian with up and down spins. With an increasing positive $g$ the relative energy difference of the up spin occupied band and the down spin unoccupied band will reduce and the bulk gap vanishes when those two bands meet. If we further increase $g$ two bands overlap and the system become ferromagnetic metal. It's the Rashba coupling that open a gap at the degenerate points of the two bands which form a surface in the three dimensional Brillouin zone and the system become a ferromagnetic insulator. From Fig 4 we see the geometry of $\Phi_{i}$s changes very little with $g$, it's a reflection that the bundle formed by the occupied bands are rather inert in the exchange field as long as the original bulk gap is still open. However, the geometry of the edge state spectrum changes a lot, the degenerate cross points are pushed away form time-reversal symmetric point and little gaps are opened at the edge. Generally the symmetry-breaking fields may affect the edge state spectrum from two different ways, one is to act directly on the edge states, the other is by changing the geometry of the bands. The direct action may change the position of degenerate points and open gaps at the degenerate point if the fields doesn't commutate with the effective edge state Hamiltonian. So even if the $\Phi_{i}$s indicate gapless edge states, gaps can still be opened by direct interaction. In this case, because the geometry structure varies very little from the $g=0$ case we show the change of the edge state spectrum comes solely from the direct interaction. We study the vector spaces spanned by the edge states and find there is almost no difference from the $g=0$ case. We also compare the edge state spectrum and that of a same system except exchange field only present at outmost three layers and they are almost identical. When the exchange field large enough the occupied bands and the unoccupied bands contact and the geometry structure changes. We see at a certain $g$ the two $\Phi_{i}$s switch partners and the occupied bands naturally decoupled two bands each with a Chern number $1$ . The system come to a quantum Hall state. In this case because of the direct interaction of the large exchange field, part of the edge state spectrum have merged into the bulk energy bands. It's no longer easy to compare the geometry of $\Phi_{i}$s and that of the edge state spectrum. However, $\Phi_{i}$s still predict two branches of edge state through the bulk gap at each edge and indicate where the new type of edges state occurs.\\

Another important example is the magnetic topological insulators in \cite{Yu2010}. Based on their first principle calculation result they predict ferromagnetic order appears in the $Bi_2Te_3$, $Bi_2Se_3$, and $Sb_2Te_3$ when doped with transition metal elements. Quantized Hall conductance can be observed in two dimensional thin film of the ferromagnetic insulator. The quantized Hall conductance can be calculated by the Kubo's formula and the change of topological structure of the energy bands are discussed in \cite{Yu2010}. Certainly we can explain the quantized Hall conductance by gapless edge state. We extend their model to a tight binding one.  When the exchange field is zero in $Bi_2Te_3$ film the band structure is just like the HgTe quantum well. So we compare the $\Phi_{i}$s and the edge state spectrum when exchange field is nonzero in Fig 6. It's easy to see how change of the geometric structure lead to the formation of the quantum Hall type of edge state. The discussion of the change topological structure of the bands in \cite{Yu2010} depend on the fact the Hamiltonian can be decoupled to two $2\times2$ blocks and two occupied bands can be easily discriminated. The topological property is obtained from the analogy with the Bernevig-Hughes-Zhang model. This property of the Hamiltonian can be easily destroyed by e.g Rashba interaction with the substrate and the eigenvectors of Hamiltonian won't natually form two bands. In our geometrical approach two bands can still be defined and the edge state properties can be analyzed in this way. For the same reason the three-dimensional edge state properties of those doped semiconductors can't be obtained from the analogy because the Hamiltonian can't be decoupled except at $k_z=0$. So we analysis the edge state property of the three-dimensional material by the geometrical property of the bands. Here we use a tight-banding model as in \cite{Li2010} with a exchange field in \cite{Yu2010} for doped $Bi_2Se_3$ and we consider the edge state in the $k_x-k_z$ plane. Without the exchange field the $\Phi_{i}$s and the edge state spectrums are just like the case in $k_x-k_y$ plane discussed above except the Dirac cone is anisotropic. When $k_z=0$ the Hamiltonian has the same form as the two-dimensional case and a large enough exchange field can produce quantum Hall type of edge states. When $k_z\neq0$ and exchange field is zero the $\Phi_{i}$s and the edge state is gaped. A large enough exchange field can also produce the gapless edge state as indicated in Fig 6. However, the farther away from $k_z=0$ the larger exchange field is needed to produce the gapless edge state. So if a fixed exchange field can produce quantum Hall type of edge state at $k_z=0$ when we go far enough from $k_z=0$ the gap will be opened as in Fig 6. At the critical $k_z=\pm k_{z0}$ the two naturally decoupled bands switch partners and become two trivial bands when $|k_z|$ increase further. At $(0,0,\pm k_{z0})$ the occupied bands and the unoccupied bands contact or ``kiss" each other and form three dimensional anisotropic Dirac points as analogys of the two-dimensional Dirac points in graphene. Here through analysis the geometrical structure of the bands we predict there is a topological-insulator to topological-semimetal phase transition when bulk $Bi_2Se_3$ is doped with transition metal elements. From the Fig 6 it can be easily inferred there is one Fermi arcs of the edge state as in \cite{Wan2011} at each of the opposite surface in the topological-semimetal phase. The bulk and surface energy spectrums can be measured by ARPES.\\

Our calculations with various tight-binding models show that the geometrical structure of bands can be a powerful tools to study the the edge state properties of insulators. The algorithm to calculate $\Phi_{i}$s is very simple. Thus we deem it should be a common procedure to calculate the geometry structure of the energy bands in studying the surface structure of insulators in the first principle energy band calculation. The geometry structure of the bands doesn't depend on the symmetry of the system and provide more edge state information than the topological invariants, so our method broadened the scope of the study of the edge state properties of the insulators.

\begin{figure}[h]
\includegraphics[width=5 in,clip=true]{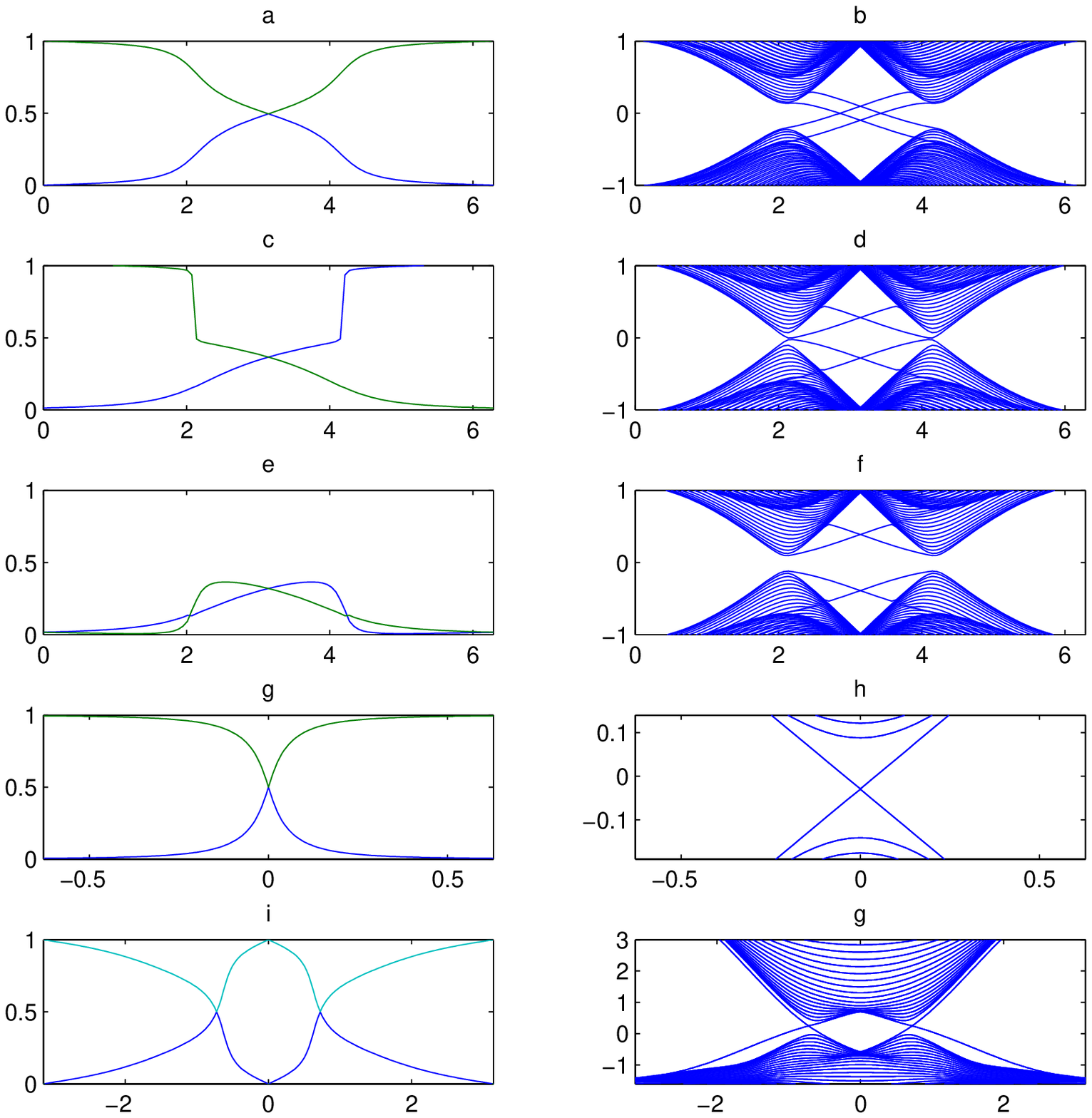}

\caption{\textbf{Comparison of the $\Phi_{i}$s and the edge state spectrum of some time reversal symmetric two dimensional insulators.} All pictures in this article are arranged as $\Phi_{i}$s at left and edge state spectrum at right. (a)-(f) show the topological and trivial insulator phase transition in graphene when $\lambda$ increases. (g)-(h) show the result in HgTe quantum well. To make the edge state more localized we use a parameter $M=-0.1$. (i)-(j) show the result in a $Z_2$ trivial insulator in \cite{Qi2006}} \label{1 }
\end{figure}

 \begin{figure}[h]
\includegraphics[width=5 in,clip=true]{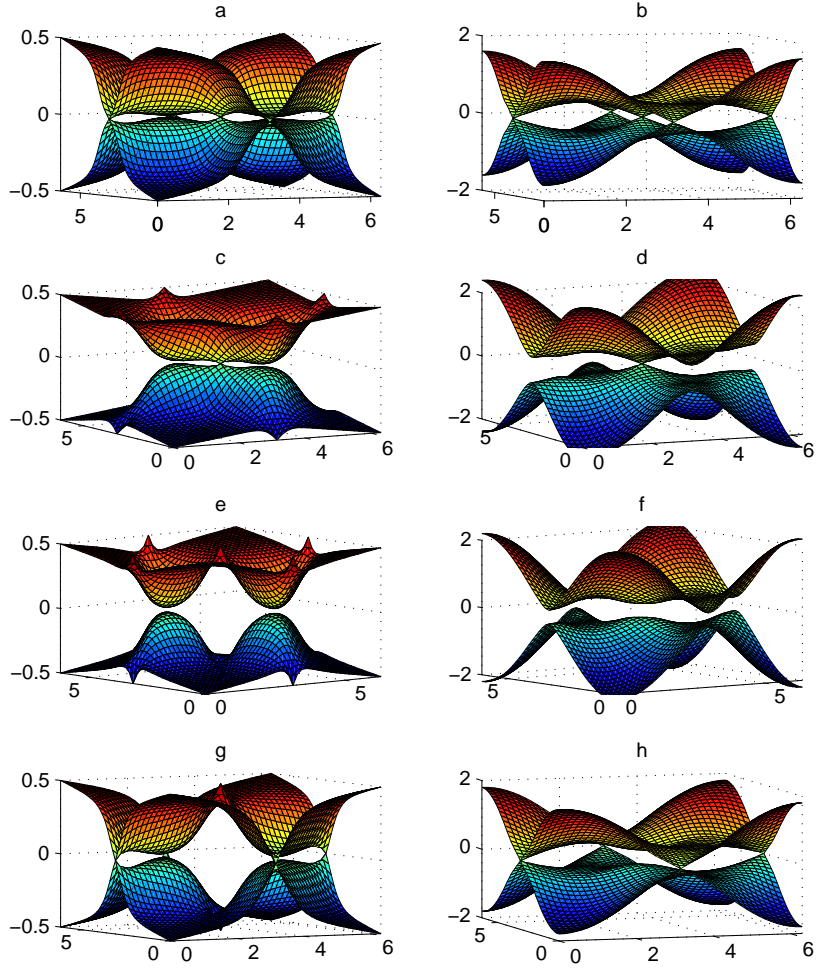}
\caption{\textbf{Comparison of the $\Phi_{i}$s and the edge state spectrum of some time reversal symmetric three dimensional insulators in \cite{Fu2007}} (a)-(b) show the $1;(111)$ case,there are three Dirac points.(c)-(d) show the $1;(1\overline{1}\overline{1})$ case, there is one Dirac point. (e)-(f) show the $1;(111)$ case there are no Dirac point. (g)-(h) show the $0;(1\overline{1}\overline{1})$ case, there are two Dirac points } \label{2}
\end{figure}

\begin{figure}[h]
\includegraphics[width=5 in,clip=true]{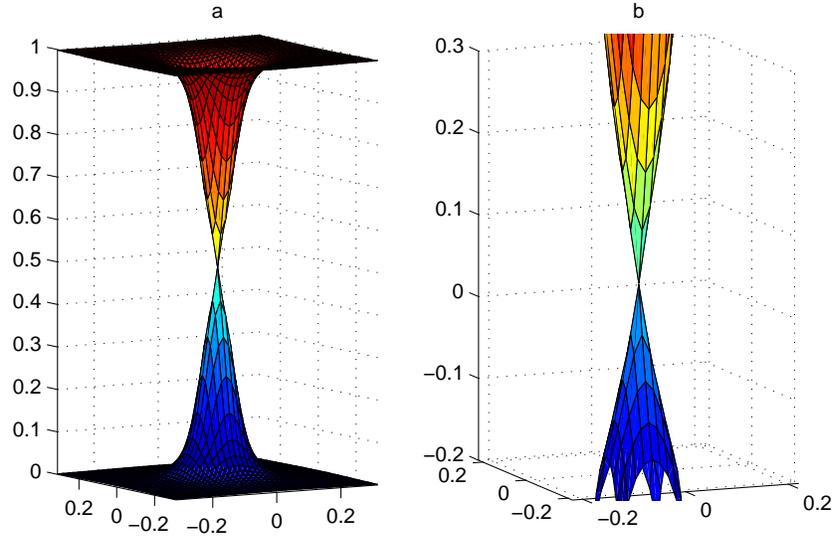}
\caption{\textbf{Comparison of the $\Phi_{i}$s and the edge state spectrum of $Bi_2Se_3$.} There is one Dirac point in this case. It's a strong topological insulator according to the classification in \cite{Fu2007}.} \label{3}
\end{figure}
\begin{figure}[h]
\includegraphics[width=5 in,clip=true]{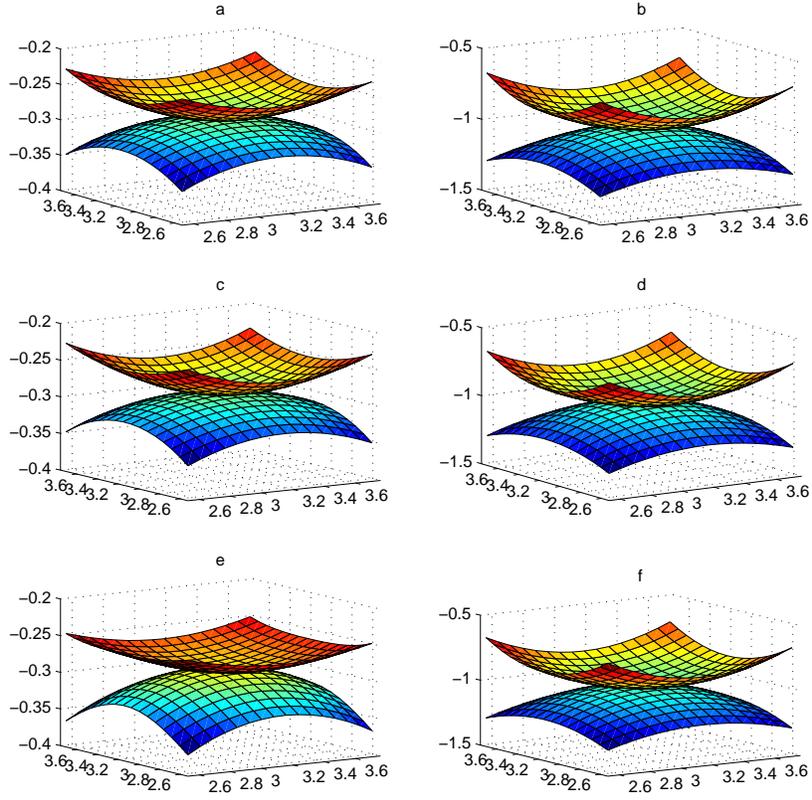}
\caption{\textbf{Comparison of the $\Phi_{i}$s and the edge state spectrum of topological crystalline insulators} (a)-(b) the case discussed in \cite{Fu2007}. (c)-(f) the $C_4$ symmetry is broken by using different $t_{2}^{'}$ in $x$ and $y$ directions. In (c)-(f) $t_{2x}^{'}=0.7$ and $t_{2y}^{'}=0.3$. In (g)-(h) $t_{2x}^{'}=1.5$ and $t_{2y}^{'}=0$  }\label{4}
\end{figure}

 \begin{figure}[h]
\includegraphics[width=5 in,clip=true]{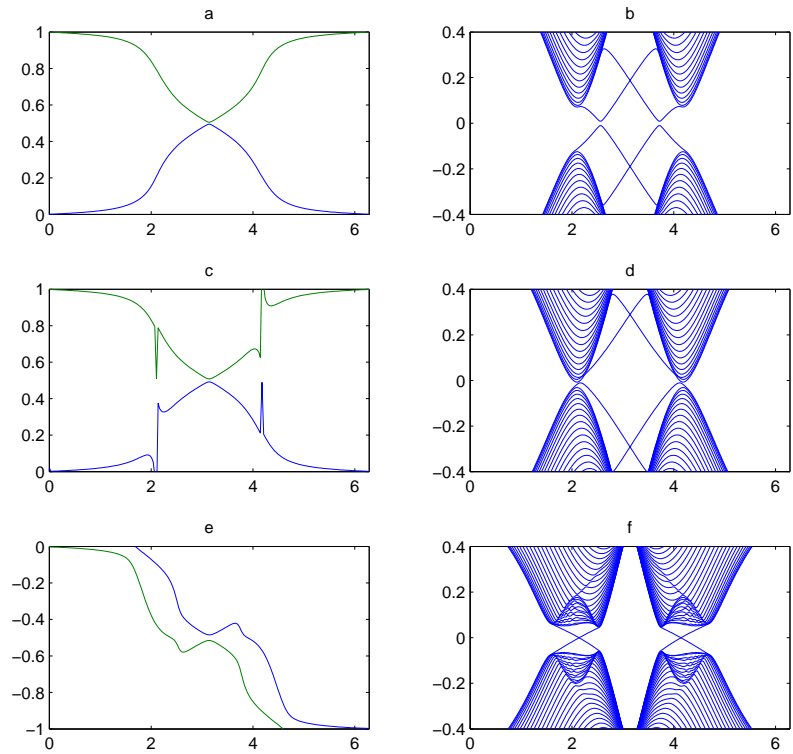}
\caption{\textbf{Comparison of the $\Phi_{i}$s and the edge state spectrum of time-reversal-symmetry-broken two dimensional topological insulators.} Here parameters of the time-reversal symmetric part of Hamiltonian in \cite{Kane2005b} are used with $\lambda_v=0$. The Hamiltonian of the exchange field is the same in \cite{Yang2011}. When the parameter $g$ increases the system experience a quantum spin Hall state to quantum Hall state phase transistion.} \label{5}
\end{figure}

\begin{figure}[h]
\includegraphics[width=2.5 in,clip=true]{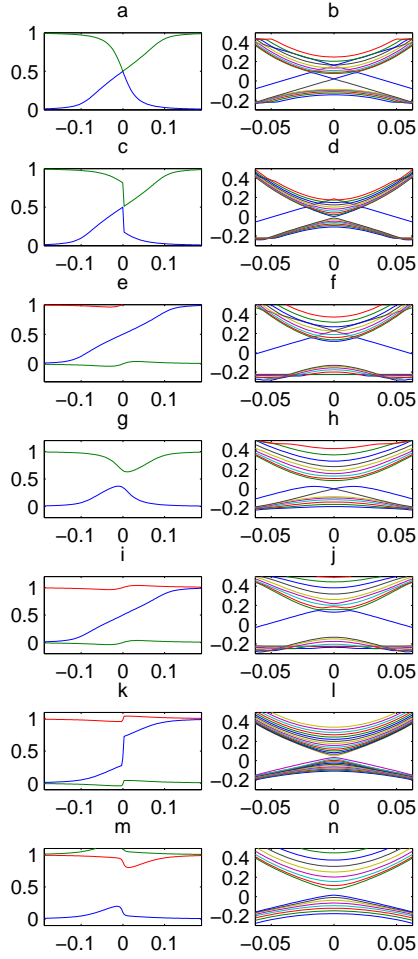}
\caption{\textbf{Comparison of the $\Phi_{i}$s and the edge state spectrum of doped $Bi_2Se_3$ film and three dimensional doped $Bi_2Se_3$ at different $k_z$} (a)-(f) corresponding to the doped $Bi_2Se_3$ film or the three dimensional doped $Bi_2Se_3$ at different $k_z=0$. With increasing exchange field parameter $g$ one of the two bands become trivial, the edge state become quantum Hall type. (g)-(j) indicated at $k_z\neq 0$ a large enough exchange field can produce quantum Hall type edge state. For a fixed exchange field as in (e) (f) (i) and (j), it's shown in (k) (l) at a certain $|k_z|$ occupied band and unoccupied band touch. (m) (n) shows when $|k_z|$ increased further the bands become trivial and there are on edge states.}\label{6}
\end{figure}

\end{document}